\documentclass[aps,prl,twocolumn,superscriptaddress,english]{revtex4}
\usepackage{amssymb}
\usepackage{graphicx}
\usepackage{amsmath}
\usepackage{soul}
\usepackage{amsthm}
\usepackage{amsfonts}
\usepackage[T1]{fontenc}
\usepackage[latin9]{inputenc}
\usepackage{array}
\usepackage{multirow}
\usepackage{color}
\usepackage{esint}
\usepackage{bm}
\usepackage{color}
\usepackage{bbm}
\usepackage{hyperref}
\usepackage{babel}
\usepackage{titlesec}

\begin{document}

\global\long\def\id{\mathbbm{1}}
\global\long\def\ui{\mathbbm{i}}
\global\long\def\ud{\mathrm{d}}

\title{Two dimensional vertex-decorated Lieb lattice with exact mobility edges and robust flat bands}
\author{Yucheng Wang}
\thanks{Corresponding author: wangyc3@sustech.edu.cn}
\affiliation{Shenzhen Institute for Quantum Science and Engineering,
Southern University of Science and Technology, Shenzhen 518055, China}
\affiliation{International Quantum Academy, Shenzhen 518048, China}
\affiliation{Guangdong Provincial Key Laboratory of Quantum Science and Engineering, Southern University of Science and Technology, Shenzhen 518055, China}
\author{Long Zhang}
\affiliation{School of Physics and Institute for Quantum Science and Engineering, Huazhong University of Science and Technology, Wuhan 430074, China}
\author{Yuhao Wan}
\affiliation{International Center for Quantum Materials, School of Physics, Peking University, Beijing 100871, China}
\author{Yu He}
\thanks{Corresponding author: hey6@sustech.edu.cn}
\affiliation{Shenzhen Institute for Quantum Science and Engineering,
Southern University of Science and Technology, Shenzhen 518055, China}
\affiliation{International Quantum Academy, Shenzhen 518048, China}
\affiliation{Guangdong Provincial Key Laboratory of Quantum Science and Engineering, Southern University of Science and Technology, Shenzhen 518055, China}
\author{Yongjian Wang}
\thanks{Corresponding author: wangyongjian@amss.ac.cn}
\affiliation{School of Mathematical Sciences, Laboratory of Mathematics and Complex Systems, MOE, Beijing Normal University, 100875 Beijing, China}
\begin{abstract}
The mobility edge (ME) that marks the energy separating extended and localized states is a most important
concept in understanding the metal-insulator transition induced by disordered or quasiperiodic potentials.
MEs have been extensively studied in three dimensional disorder systems and one-dimensional quasiperiodic systems. However, the studies of MEs in two dimensional (2D) systems are rare. Here we propose a class of 2D vertex-decorated Lieb lattice models with quasiperiodic potentials only acting on the vertices of the Lieb lattice or extended Lieb lattices. By mapping these models to the 2D Aubry-Andr\'{e} model, we obtain exact expressions of MEs and the localization lengths of localized states,
and further demonstrate that the flat bands remain unaffected by the quasiperiodic potentials. Finally, we propose a highly feasible scheme to experimentally realize our model in a quantum dot array. Our results open the door to studying and realizing exact MEs and robust flat bands in 2D systems.
\end{abstract}
\maketitle

{\em Introduction.---} Quantum interference in disordered systems can completely suppress the diffusion of particles, which is a fundamental phenomenon known as Anderson localization~\cite{Anderson1958,RMP2,Kramer,RMP1}. In three-dimensional (3D) systems,
the metal-insulator Anderson transition (AT) can occur as a function of disorder strength or energy, and the latter produces the mobility edges (MEs), which mark the critical energies separating localized eigenstates from extended ones~\cite{RMP2,Kramer,RMP1,Lagendijk2009}.
Localization phenomena are sensitive to the spatial dimensionality of a system.
According to the one-parameter scaling theory~\cite{Thouless,Anderson1979}, in conventional cases, AT and MEs exist in 3D systems, but are absent in one-dimensional (1D) and two-dimensional (2D) systems.
Nevertheless, it is not easy to introduce microscopic models in 3D systems to understand the physical mechanisms of
the MEs, so it is highly important to develop models with MEs and explore the conditions that give rise to MEs in lower dimensions. Several physical mechanisms that can create MEs were uncovered, for example by introducing a magnetic field, the spin-orbit coupling, or interparticle interactions in lower dimensional systems~\cite{Aspect2007,Izrailev2012,Lee1981,Su2016,Punnoose2005,PSheng,Fastenrath,SSSha,Jia2022}, in which introducing quasiperiodic potentials in place of random disorders has been studied most extensively.

Quasiperiodicity is the middle ground between periodicity and disorder, and the AT and MEs can exist even in 1D quasiperiodic systems, which have attracted great attentions in both
theory~\cite{AA,Xie1988,Biddle2009,Biddle,Ganeshan2015,Danieli2015,XDeng,XLi,HYao,Wang1,Wang2020,XPLi,TongLiu,WangLiu,Ribeiro} and experiment~\cite{Roati,Bloch1,Gadway1,Gadway2,WeiYi2022,Shimasaki2022,HaoLi2022}. Because 1D systems allow the application of several interesting analytical methods, such as the dual transformation~\cite{AA} and global theory~\cite{Avila}, the AT transition point and the position of MEs can be precisely determined in some special cases. The studies of 1D quasiperiodic systems provide numerous models with MEs, which show the abundant mechanisms to induce MEs. For example, the MEs can be obtained by introducing the next-nearest-neighbor, exponential or power-law hopping terms or a spin-orbit coupling term to a quasiperiodic system. Moreover, these models with MEs, especially with exact MEs provide a solid foundation for studying abundant localization phenomena in 1D systems~\cite{XiaopengLi2015,Modak2015,Wei2019,Hsu2018,Balachandran2019,Yin2020,Yanxia2020,Kulkarni2017}. We emphasize that although most of the studies of localization phenomena are based on a concrete model, it is generally accepted that the obtained conclusions are widely suitable for other 1D systems with MEs.

Compared with 1D systems, 2D materials and devices are more widespread, and two dimension is the marginal dimension for localization~\cite{Anderson1979,PSheng,Fastenrath,SSSha,White2020}. Thus, the studies of 2D AT and MEs are undoubtedly important for both the fundamental physics and potential applications. However, the studies of AT and MEs in 2D quasiperiodic systems are just underway~\cite{Bordia2017,Huang2019,Rossignolo2019,Gautier2021,Pupillo,Schneider,Zhihao,Castelnovo}, and 2D models with exact MEs are rare, which leads to that the study of the localization physics in 2D systems is rootless and the physical mechanisms inducing MEs in 2D systems are still vague. For 2D system, the localization length-scale near a transition point is usually too large to be unambiguously numerically calculated and observed in an experiment~\cite{White2020}. Thus, analytical results of AT or MEs are especially important to {\color{blue} study} the localization physics of 2D systems.
Nevertheless, the analytical methods in 1D systems are difficult to be directly extended to 2D cases, so new ways and models need to be introduced to obtain exact expressions of MEs.

\begin{figure}
\hspace*{-0.15cm}
\centering
\includegraphics[width=0.5\textwidth]{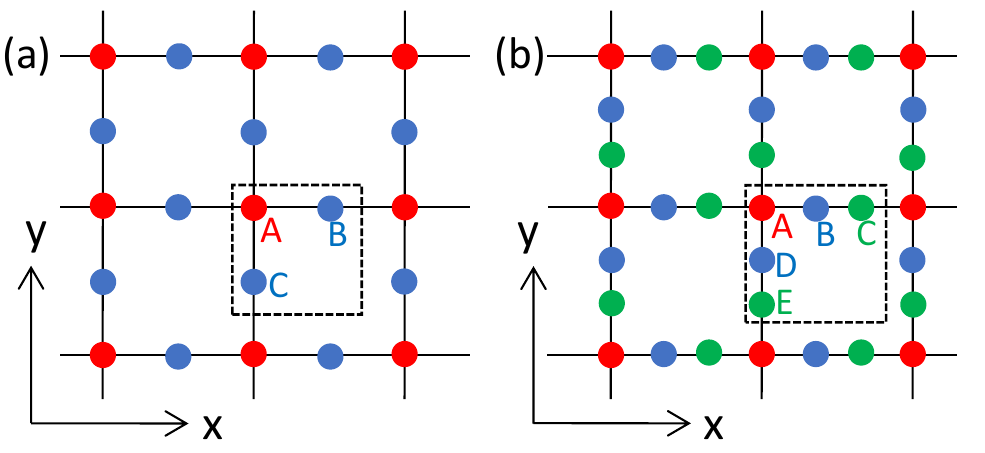}
\caption{\label{01}
Schematic figures of (a) the Lieb lattice and (b) an extended Lieb lattice models representing the edge-centered square lattice with three (A, B and C in (a)) and five lattice sites (A-E in (b)) per unit cell, respectively. The quasiperiodic potentials only act on the vertices (red spheres).}
\end{figure}

In this work, we propose a class of 2D vertex-decorated Lieb lattice (VDLL) models with exact MEs, where quasiperiodic potentials are inlaid in the Lieb lattice or extended Lieb lattices with equally spaced sites and only act on the vertices (red spheres in Fig.~\ref{01}).  Lieb lattice is one of the most popular in the family of flat-band models, which has three lattice sites per unit cell~\cite{Lieb1989} [Fig.~\ref{01}(a)], or as an extended version, has five [Fig.~\ref{01}(b)] or more sites per unit cell~\cite{DZhang2017,WJiang2019,Pal2019,XMao2020}. This model has been used to explore various interesting physics~\cite{Noda2009,Goldman2011,Julku2016,Ozawa2017,Whittaker,Schmelcher2019,Ma2020,FLiu2022}, and realized with photonic~\cite{Mukherjee2015,Vicencio2015,Diebel2016,SXia2018,SXia2016}, atomic~\cite{Taie2015,Baboux2016} and electronic systems~\cite{Drost2017,Slot2017}. We obtain the exact expressions of MEs analytically by mapping VDLL models to 2D Aubry-Andr\'{e} (AA) model and numerically by calculating the fractal dimension. The flat bands are unaffected by the quasiperiodic potentials. We further propose a novel scheme to realize the VDLL model in a 2D quantum dot system.

{\em Model and results.---} We propose a class of 2D VDLL models described by
\begin{equation}\label{Hsum}
H=\sum_{\langle ij;i'j'\rangle}(Jc^{\dagger}_{ij}c_{i'j'}+h.c.)+\sum_{ij}V_{ij}n_{ij},
\end{equation}
with $V_{ij}=$
\begin{equation}\notag
\begin{cases}
2V[\cos(2\pi\alpha_1 i+\theta_1)+\cos(2\pi\alpha_2 j+\theta_2)], \ (i,j)=(m\rho,n\rho)  \\
0, \ \textrm{otherwise},
\end{cases}
\end{equation}
where $\alpha_1$ and $\alpha_2$ are irrational numbers, $c_{ij}$($c^{\dagger}_{ij}$) is the annihilation (creation) operator that acts on site $(i,j)$, $n_{ij}=c^{\dagger}_{ij}c_{ij}$ is the particle number operator, $J$ represents the hopping strength between neighboring sites, $V$ is the quasiperiodic potential amplitude. Without loss of generality, we set $\alpha_1=\frac{\sqrt{5}-1}{2}$, $\alpha_2=\frac{\sqrt{2}}{2}$, the phase shifts $\theta_1=\theta_2=0$ and take periodic boundary conditions unless otherwise stated. The interval between the nearest neighbor vertices is set as $\rho$ in both $x$ and $y$ directions, so Fig.~\ref{01}(a) and (b) correspond to $\rho=2$ and $\rho=3$, respectively. The symbols $m$ and $n$ represent the $m-$th and $n-$th unit cell in $x$ and $y$ directions, respectively. For convenience, we set $m\ (n)=0, 1, 2, \dots, L_x-1\ (L_y-1)$ with $L_x$ ($L_y$) being the cell number in $x$ ($y$) direction in the absence of quasiperiodic potentials, and the indexes $i$ and $j$ start from $0$ to ensure that the quasiperiodic potentials only act on the vertices [the red spheres in Fig.~\ref{01}].

\begin{figure}
\hspace*{-0.1cm}
\centering
\includegraphics[width=0.5\textwidth]{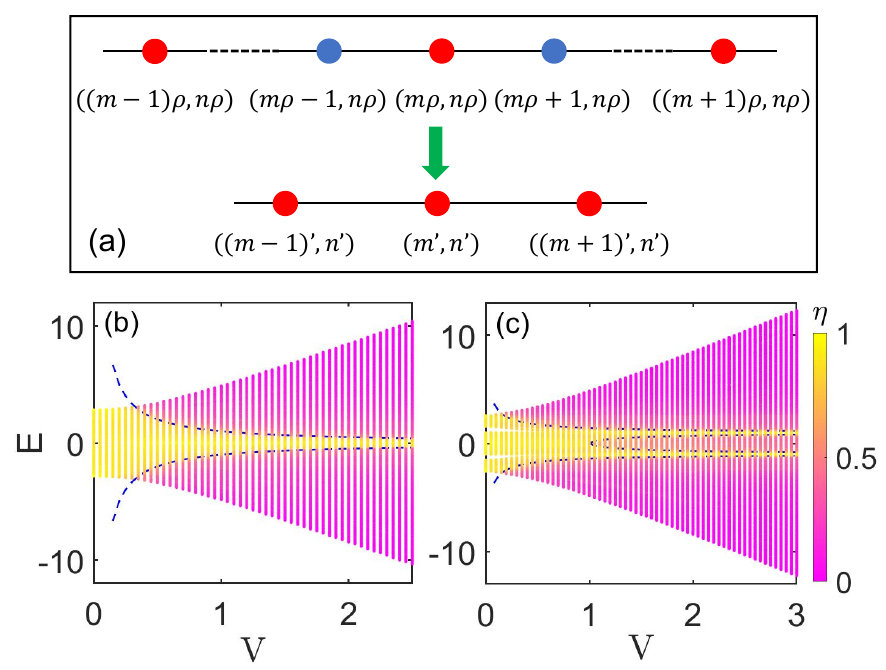}
\caption{\label{02}
(a) Mapping 2D VDLL models to 2D AA model. Here we show the mapping in $x$ direction, and the mapping is similar in $y$ direction.  Fractal dimension $\eta$ as a function of the quasiperiodic potential strength $V$ and eigenvalues for (b) $\rho=2$ and (c) $\rho=3$ with the system size $L_x=L_y=34$. The blue dashed lines represent the MEs given by Eq.~(\ref{rho2}) and Eq.~(\ref{rho3}), respectively. Here we set $J=1$ as the unit of energy.}
\end{figure}

The exact MEs and localization length can be obtained by deforming the 2D VDLL models to the 2D AA model. Suppose that an eigenstate is given by $|\psi\rangle=\sum_{i,j}u_{i,j}c^{\dagger}_{i,j}|0\rangle$, the eigenvalue equation $H|\psi\rangle=E|\psi\rangle$ leads to the following equation:
\begin{eqnarray}
E u_{m\rho,n\rho}= Ju_{m\rho+1,n\rho}+Ju_{m\rho-1,n\rho}+Ju_{m\rho,n\rho+1}\nonumber\\
+Ju_{m\rho,n\rho-1}+V_{m\rho,n\rho}u_{m\rho,n\rho}.\qquad \qquad \qquad
\label{Lambda6}
\end{eqnarray}
Here $u_{m\rho+1,n\rho}$ can be replaced by $u_{(m+1)\rho,n\rho}$ based on the transfer matrix form,
\begin{equation}
\left(
\begin{array}{c}
u_{(m+1)\rho,n\rho} \\
u_{(m+1)\rho-1,n\rho}
\end{array}
\right)=T^{\rho-1}\left(
\begin{array}{c}
u_{m\rho+1,n\rho} \\
u_{m\rho,n\rho}
\end{array}
\right)
\notag
\end{equation}
where the transfer matrix
\begin{equation*}
T^{\rho-1}=\left(
\begin{array}{cc}
  E/J& -1 \\
  1 & 0 \\
\end{array}
\right)^{\rho-1} =\left(
\begin{array}{cc}
  F_{\rho}& -F_{\rho-1} \\
 F_{\rho-1} & -F_{\rho-2} \\
\end{array}
\right)
\end{equation*}
with
\begin{eqnarray}\label{Prho}
F_{\rho}=
       \frac{1}{\sqrt{\epsilon^2-4}}\left[(\frac{\epsilon+ \sqrt{\epsilon^2-4}}{2})^{\rho}- (\frac{\epsilon- \sqrt{\epsilon^2-4}}{2})^{\rho}\right],
\end{eqnarray}
where $\epsilon=E/J$. Then we obtain
$u_{m\rho+1,n\rho}=\frac{1}{F_{\rho}}u_{(m+1)\rho,n\rho}+\frac{F_{\rho-1}}{F_{\rho}}u_{m\rho,n\rho}$. Similarly,
$u_{m\rho-1,n\rho}$ and $u_{m\rho,n\rho\pm1}$ can be replaced by $u_{(m-1)\rho,n\rho}$ and $u_{m\rho,(n\pm1)\rho}$, respectively. Substituting these results into Eq.~(\ref{Lambda6}) yields
\begin{eqnarray}
(F_{\rho}E-4JF_{\rho-1})u_{m\rho,n\rho}= Ju_{(m+1)\rho,n\rho}+Ju_{(m-1)\rho,n\rho}\nonumber\\
+Ju_{m\rho,(n+1)\rho}+Ju_{m\rho,(n-1)\rho}+F_{\rho}V_{m\rho,n\rho}u_{m\rho,n\rho}.
\label{Lambda8}
\end{eqnarray}
The indexes are divided by $\rho$, i.e., $m\rho\rightarrow m', n\rho\rightarrow n', (m\pm1)\rho\rightarrow (m\pm1)', (n\pm1)\rho\rightarrow (n\pm1)'$ and $F_{\rho}E-4JF_{\rho-1}\rightarrow E'$, as shown in Fig.~\ref{02}(a).  Then the above equation have the same form as the isotropic 2D AA model~\cite{AA,Bordia2017} with the effective quasiperiodic potential $V'_{m',n'}=F_{\rho}V_{m',n'}=2V'[\cos(2\pi\alpha_1m')+\cos(2\pi\alpha_2 n')]$, where $V'=F_{\rho}V$. The extended-localized transition point of the isotropic 2D AA model is at $|V'/J|=1$, which can be analytically obtained by using the dual transformation~\cite{SM}.
Thus the extended-localized transition points of the 2D VDLL models satisfy
\begin{equation}\label{trans}
|V'/J|=1\rightarrow |F_{\rho}V/J|=1.
\end{equation}
Since $|V'/J|>1$ ($|V'/J|<1$) corresponds to the localized (extended) states of AA models, $|F_{\rho}V/J|>1$ and $|F_{\rho}V/J|<1$ respectively correspond to the localized and extended states of the VDLL models. Here $F_{\rho}$ shown in Eq.~(\ref{Prho}) depend on energies. Thus, Eq.~(\ref{trans}) is the expression of MEs, which also applies to 3D quasiperiodic Lieb lattices~\cite{JLiu2020} with quasiperiodic potentials being added at the cross points, because these models can be mapped to a 3D AA model~\cite{Devakul2017} when the same process is also applied to the $z$ direction.  From Eq.~(\ref{Prho}) and Eq.~(\ref{trans}), when $\rho=2$, there are two MEs, given by
\begin{equation}\label{rho2}
E_c=\pm\frac{J^2}{V}.
\end{equation}
For $\rho=3$, four MEs emerge, which read
\begin{equation}\label{rho3}
E_c=\pm J\sqrt{1\pm\frac{J}{V}}.
\end{equation}

The analytical results can be numerically verified by calculating the fractal dimension that is defined as $\eta=-\lim_{N\rightarrow\infty}\ln(IPR)/\ln N$, where $IPR=\sum_{ij}u^4_{ij}$ is the inverse participation ratio~\cite{RMP1} and $N=(2\rho-1)\times L_x\times L_y$ is the number of the total lattice sites. The fractal dimension tends to $0$ and $1$ for the localized and extended states, respectively. Fig.~\ref{02}~(b) and (c) show $\eta$ of different eigenstates
as the function of $V$ and the corresponding eigenvalues $E$ for $\rho=2$ and $\rho=3$, respectively, which show that the states in the pink and yellow regions are respectively localized and extended, and they are separated by the blue dashed lines, which represent the
MEs described by Eq.~(\ref{rho2}) and Eq.~(\ref{rho3}). As expected from
the analytical results, $\eta$ suddenly changes when energies across the dashed lines. Further, for any $\rho$, one can obtain $2(\rho-1)$ MEs described by Eqs.~(\ref{Prho}) and (\ref{trans}).

Taking the advantage of the above mapping, we can also obtain the localization length. It is well known that the localization length of a AA model is $\xi=1/\ln(V'/J)$, so the localization lengths of these models in both $x$ and $y$ directions are
\begin{equation}\label{loc}
\xi(E)=\frac{\rho}{\ln|F_{\rho}V/J|}.
\end{equation}
Here the $\rho$ in the numerator originates from that the system size enlarge $\rho$ times when mapping the 2D AA model back to these models we considered. One can define a critical exponent $\nu$ by $\xi\sim(V-V_c)^{-\nu}$ or $\xi\sim(E-E_c)^{-\nu}$ and determine $\nu=1$ according to Eq.~(\ref{loc})~\cite{explainloc}, which is a general result for 2D quasiperiodic systems.

\begin{figure}
\hspace*{-0.1cm}
\centering
\includegraphics[width=0.48\textwidth]{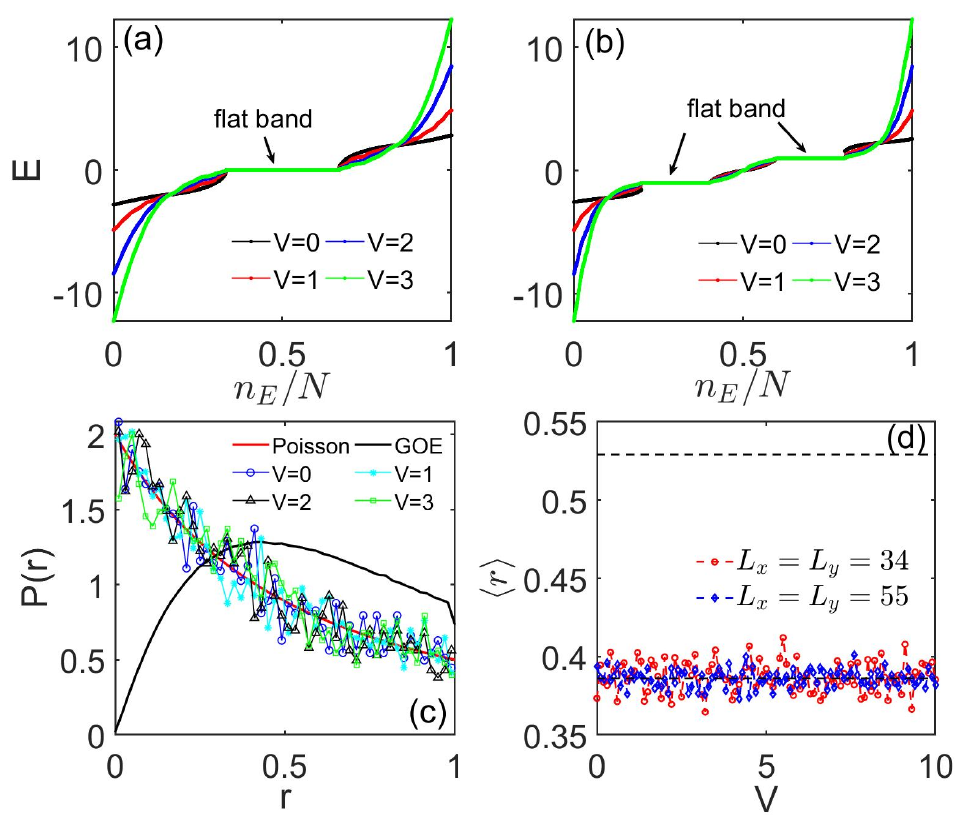}
\caption{\label{03}
 Eigenvalues as a function of $n_E/N$ with $n_E$ being the index of eigen-energies for (a) the $\rho=2$ case and (b) the $\rho=3$ case. (c) The probability distribution of the level-spacing ratio $r$ for different quasiperiodic potential strengths $V$. The system sizes in (a), (b) and (c) are $L_x=L_y=34$. (d) $\langle r\rangle$ versus $V$ for different sizes. Here we set $J=1$. }
\end{figure}

{\em Robust flat bands.---}
While flat bands are usually very fragile and easily destroyed by weak disorder or quasiperiodic potentials~\cite{FLiu2022,Goda2006,Nishino2007,Chalker2010,Bodyfelt2014,Leykam2017,Shukla2018,Roy2020,Ahmed2022,Lee2022}, here we see that the flat bands are in the extended regions, indicating that the flat bands may be not affected by the Anderson localization induced by the quasiperiodic potential. We below shall show that the flat bands of our VDLL models  are immune to the quasiperiodic or disorder potentials which act on the vertices.  Fig.~\ref{03}(a) and (b) show the eigenvalues corresponding to $\rho=2$ and $\rho=3$, respectively. When increasing the quasiperiodic potential strength, the number of eigenstates in flat bands remains unchanged, suggesting that the flat bands may be unaffected.

We then investigate the localized eigenstates in a flat band, namely the compact localized states.
To study the effect of the quasiperiodic potentials, we consider the statistical properties of the energy levels in the flat bands by calculating the level-spacing ratio $r_k=\frac{min(\delta_k,\delta_{k+1})}{max(\delta_k,\delta_{k+1})}$~\cite{Shore1993,Oganesyan2007}, where $\delta_k=E_{k+1}-E_k$ is the energy spacing. Here eigenvalues $E_k$ have been listed in ascending order, and $k\in(\frac{N}{3},\frac{2N}{3})$ for the $\rho=2$ case. In the localized region, the spectral statistics are Poisson, which yields the probability distribution of the level-spacing ratio $r$: $P(r)=\frac{2}{(1+r)^2}$ [red line in Fig.~\ref{03}(c)], and its mean value $\langle r\rangle\approx 0.387$~\cite{Oganesyan2007}. In the extended region, the spectral statistics follow Gaussian-orthogonal ensemble yielding $\langle r\rangle\approx 0.529$, and the distribution of the ratio $r$ can be obtained by using random matrices~\cite{Oganesyan2007,explain2} [black line in Fig.~\ref{03}(c)]. Fig.~\ref{03}(c) and (d) display the distribution of $r$ and the averaged $r$~\cite{explain3,explain4}, respectively.
As expected, despite changing the quasiperiodic potential strength, the level statistics remain Poissonian, implying that all states in the flat band remain unaffected. Further, if replacing the quasiperiodic potentials with random disorder ones, the flat bands are not affected either.

These results can be understood from Eq.~(\ref{Prho}) and Eq.~(\ref{Lambda8}). From Eq.~(\ref{Lambda8}), the effective potentials in the 2D AA model are $F_{\rho}V_{m',n'}$. From Eq.~(\ref{Prho}), $F_{\rho}=E/J$ when $\rho=2$ and $F_{\rho}=(E/J)^2-1$ when $\rho=3$. For the $\rho=2$ case, the flat band
is at the energy $E/J=0$, and when $\rho=3$, the flat bands correspond to $E/J=\pm 1$. Thus, at the flat bands, we have $F_{\rho}=0$, which leads to
$F_{\rho}V_{m',n'}=0$ regardless of what form of $V_{m',n'}$. In the supplementary materials~\cite{SM}, we further discuss the underlying mechanism for the occurrence of the robust flat bands.
To conclude, flat bands are very easily destroyed when disorder or quasiperiodic potentials act on all sites~\cite{SM}, but when these potentials just act on the vertices of the Lieb lattices, the flat bands are unaffected.

\begin{figure}
\hspace*{-0.1cm}
\centering
\includegraphics[width=0.47\textwidth]{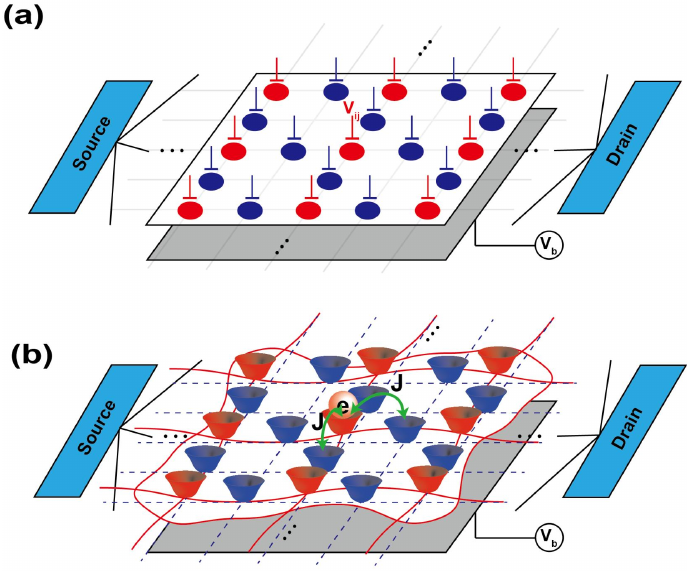}
\caption{\label{04}
A proposed experimental system using a 2D quantum dot array. (a) Sketch of a quantum dot array. The potential offsets of the red dot are adjusted by their plunger gates, and the blue dot potentials are fixed. The dot array is connected by source and drain leads on both edges via multiplexing wires or directly tunnel coupled. (b) Electron transport in a modulated Lieb lattice array. The system is operated at the one electron regime, to avoid unwanted multiple electron interactions. The bottom gate $V_b$ is used to tune the global offset to ensure only a single electron is loaded onto this system~\cite{Kiczynski2022}. The hopping $J$ is fixed between all the nearest-neighbor sites and is engineered by controlling the distances between the dots. }
\end{figure}

{\em Experimental realization.---}
Due to the realization of the Lieb lattice in many experiments~\cite{Mukherjee2015,Vicencio2015,Diebel2016,SXia2018,SXia2016,Taie2015,Baboux2016,Drost2017,Slot2017},
our proposed VDLL model is of high feasibility and can be simulated in several different systems.
Here we take the quantum dot system as an illustration and present a concrete realization proposal.
Our designed 2D quantum dot system can be either a gate defined dot array~\cite{Dehollain2020,Hendrickx2021Four,Philips2022Six,Kandel2019Four,Menno20172D,Mills2019,Takeda2022Three} or a scanning tunnelling microscope atomic precision lithographed dot array~\cite{Kiczynski2022,Hill20152D}, as shown in Fig.~\ref{04}(a). The 2D array is connected to a source and drain gates on both edges via multiplexed leads or directly tunnel coupled, and the transport conductance signal can be obtained by taking the current differential from source to drain. The dot array consists of two types of quantum dots, labeled red and blue.
Besides the necessary gates to define the quantum dots (which is not shown in Fig.~\ref{04}), each dot has a plunger gate to tune its chemical potential. Experimentally, all the possible initial potential offsets could be aligned by voltages applied on those corresponding plunger gates~\cite{Dehollain2020,Kandel2019Four,THensgens2017,HQiao2020}. Next, the potential offsets of red dots are individually tunned by voltages on the vertical gates for corresponding values $V_{ij}=2V[\cos(2\pi \alpha_1 i+\theta_1)+\cos(2\pi \alpha_2 j+\theta_2)]$, while the potentials of the blue ones are fixed, where the virtual gate method could be used to remove signal couplings between the neighboring gates~\cite{Mills2019,CVolk2019}. The nearest-neighbor hopping $J$ is fixed between the sites. As shown in Fig.~\ref{04}(b), the system is operated at one electron regime by tuning the global bottom gate $V_b$, hence the onsite energy and intersite Coulombic interaction terms of Hubbard Hamiltonian~\cite{Kiczynski2022} are not functioning. Therefore, under zero-magnetic field condition, the spinless system is depicted by the Hamiltonian~\eqref{Hsum}.

Here we give a simple check of the feasibility of this system. Taking gate operation voltages within $\sim$1 volt and considering a gate lever arm of $\sim$0.03, the potential energy of a single dot is fully tunable within a region of 30 meV, corresponding to a relative tunning range from -15 meV to 15 meV. Usually, the tunneling rate $J$ can be easily engineered from $\sim\mu$eV to $\sim$meV level; here we pick $J=150$ $\mu$eV for example. Thus $V/J$ can be tuned from -100 to 100, which fully covers the parameter range to observe the predicted phenomena in Fig.~\ref{02}(b). In supplementary materials~\cite{SM}, we also show that although the quasiperiodic potential $V_{ij}$ is replaced by $V_{ij}+0.2V_{ij}R_{ij}$ with $R_{ij}\in[-1,1]$ being random numbers, the MEs are scarcely influenced. Thus, the experimental realization of $V_{ij}$ is of high fault tolerance.

The predicted MEs could be investigated based on the transport signal~\cite{SM}. For Fermi
surface in the localized regions, the conductivity decays exponentially with system sizes, while in the extended regions, the conductivity is independent of the system sizes. Since the conductivity attenuates rapidly with sizes in localized regions, it suddenly
changes when energies across the MEs when the system size is large enough~\cite{SM}. Thus, one can detect MEs by detecting the scaling behaviors
of the conductance or detecting the conductance in a large system.

{\em Discussion and conclusion.---} We have proposed a class of 2D VDLL models, where quasiperiodic potentials only act on the vertices of the Lieb lattice and extended Lieb lattices, and derived the expressions of MEs and localization lengths by deforming these models to the 2D AA model. The exact MEs are further numerically verified by calculating the fractal dimension. We further found that the flat bands remain unaffected by such added quasiperiodic potentials. Finally, we studied in detail the experimental realization of a VDLL model based on a 2D quantum dot array.  Our work opens the door to searching for exact MEs and robust flat bands in 2D systems.

In supplementary materials~\cite{SM}, we also consider the case that quasiperiodic potentials only act on the edge sites [blue spheres in Fig.~\ref{01}(a)], and find the existence of the critical zone. Thus, there are also the MEs separating the extended or localized states from critical ones in 2D systems. It can be seen that the quasiperiodic or disorder potentials acting on different types of elements or lattice sites of a 2D system may produce different rich physics phenomena, which will motivate the construction of interesting models and the discovery of new physical phenomena.

\begin{acknowledgments}
We thank X.-J. Liu and P. Huang for valuable discussions. This work is supported by the National Key R\&D
Program of China under Grant No.2022YFA1405800, the National Natural Science Foundation of China (Grants No.12104205, 62174076, 92165210),  the Key-Area Research and Development Program of Guangdong Province (Grant
No. 2018B030326001), Guangdong Provincial Key Laboratory (Grant No.2019B121203002).
L. Z. acknowledges support from the startup grant of Huazhong University of Science and Technology (Grant No. 3004012191). Y. H. is supported by the Shenzhen Science and Technology Program (Grant No. KQTD20200820113010023). Y. W. is supported by NSFC (Grant 12061031).
\end{acknowledgments}



\clearpage
\linespread{1}

\global\long\def\id{\mathbbm{1}}
\global\long\def\ui{\mathbbm{i}}
\global\long\def\ud{\mathrm{d}}

\setcounter{equation}{0} \setcounter{figure}{0}
\setcounter{table}{0} 
\renewcommand{\theparagraph}{\bf}
\renewcommand{\thefigure}{S\arabic{figure}}
\renewcommand{\theequation}{S\arabic{equation}}

\onecolumngrid
\flushbottom
\section*{\large Supplementary Material:\\Two dimensional vertex-decorated Lieb lattice with exact mobility edges and robust flat bands}
In the Supplementary Materials, we first give a brief introduction to two dimensional (2D) Aubry-Andr\'{e} (AA) model and give a concrete example to show the process of mapping the 2D vertex-decorated Lieb lattice (VDLL) models to 2D AA model. Then, we uncover the underlying mechanism for the occurrence of the robust flat bands, and study the conduction and the error influence in the experimental realization. Finally, we discuss the two cases that quasiperiodic potentials only act on the edge sites and all Lieb lattice sites.

For convenience, we rewrite the Hamiltonian:
\begin{equation}\label{Hsums}
H=\sum_{\langle ij;i'j'\rangle}(Jc^{\dagger}_{ij}c_{i'j'}+h.c.)+\sum_{ij}V_{ij}n_{ij},
\end{equation}
with
\begin{equation}\notag
V_{ij}=
\begin{cases}
2V[\cos(2\pi\alpha_1 i+\theta_1)+\cos(2\pi\alpha_2 j+\theta_2)],\ \  (i,j)=(m\rho,n\rho)  \\
0, \ \textrm{otherwise},
\end{cases}
\end{equation}
We set $J=1$ and $\theta_1=\theta_2=0$ in the following discussions.

\section{I. A brief introduction to 2D AA model}
The generalization of the 1D AA model to $d$ dimension is described by the Hamiltonian~\cite{2DAA1}: $H=\sum_{\vec{r}}\sum_{j=1}^{d}(c^{\dagger}_{\vec{r}}c_{\vec{r}+\hat{u}_j}+H.c.)+
\sum_{\vec{r}}V(\vec{r})c^{\dagger}_{\vec{r}}c_{\vec{r}}$, with $V(\vec{r})=2V\sum_{j=1}^{d}\cos(2\pi\vec{b}_j\cdot\vec{r}+\phi_j)$. In this section, we consider $d=2$, i.e., 2D AA model, and set $\phi_1=\phi_2=0$, so the Hamiltonian reduces to Eq.~\ref{Hsums} with $V_{ij}=2V[\cos(2\pi\alpha_1i)+\cos(2\pi\alpha_2j)]$. The 2D AA model has been realized in experiment~\cite{2DAA2}.
Suppose that an eigenstate is described by $|\psi\rangle=\sum_{j,k}u_{j,k}c^{\dagger}_{j,k}|0\rangle$, the eigenvalue equation $H|\psi\rangle=E|\psi\rangle$ gives:
\begin{equation}
E u_{j,k}= J(u_{j+1,k}+u_{j-1,k}+u_{j,k+1}+u_{j,k-1})+2V[\cos(2\pi\alpha_1j)+\cos(2\pi\alpha_2k)]u_{j,k}.
\label{LambS}
\end{equation}
By using the dual transformation
 $u_{j,k}=\frac{1}{L}\sum_{n,m}v_{n,m}e^{-i(2\pi n\alpha_1 j+2\pi m\alpha_2 k)}$,
Eq.~\ref{LambS} becomes
\begin{equation}
E v_{n,m}= V(v_{n+1,m}+v_{n-1,m}+v_{n,m+1}+v_{n,m-1})+2J[\cos(2\pi\alpha_1n)+\cos(2\pi\alpha_2m)]v_{n,m}.
\label{LambSk}
\end{equation}
Eq.~\ref{LambSk} is self-dual to the original Hamiltonian defined in Eq.~\ref{LambS} when $V=J$. Thus, the extended-localized transition point is at $V=J$, and no MEs exist.

The 2D AA model discussed above is isotropic. If the hopping strength or quasiperiodic potential strength is unequal in $x$ and $y$ directions, the system will show richer phenomena. For example, there may exist the wavefunction that is extended in one direction but localized in the other direction, namely that the particle can move only in a single direction.

\begin{figure}[t]
\includegraphics[width=0.6\textwidth]{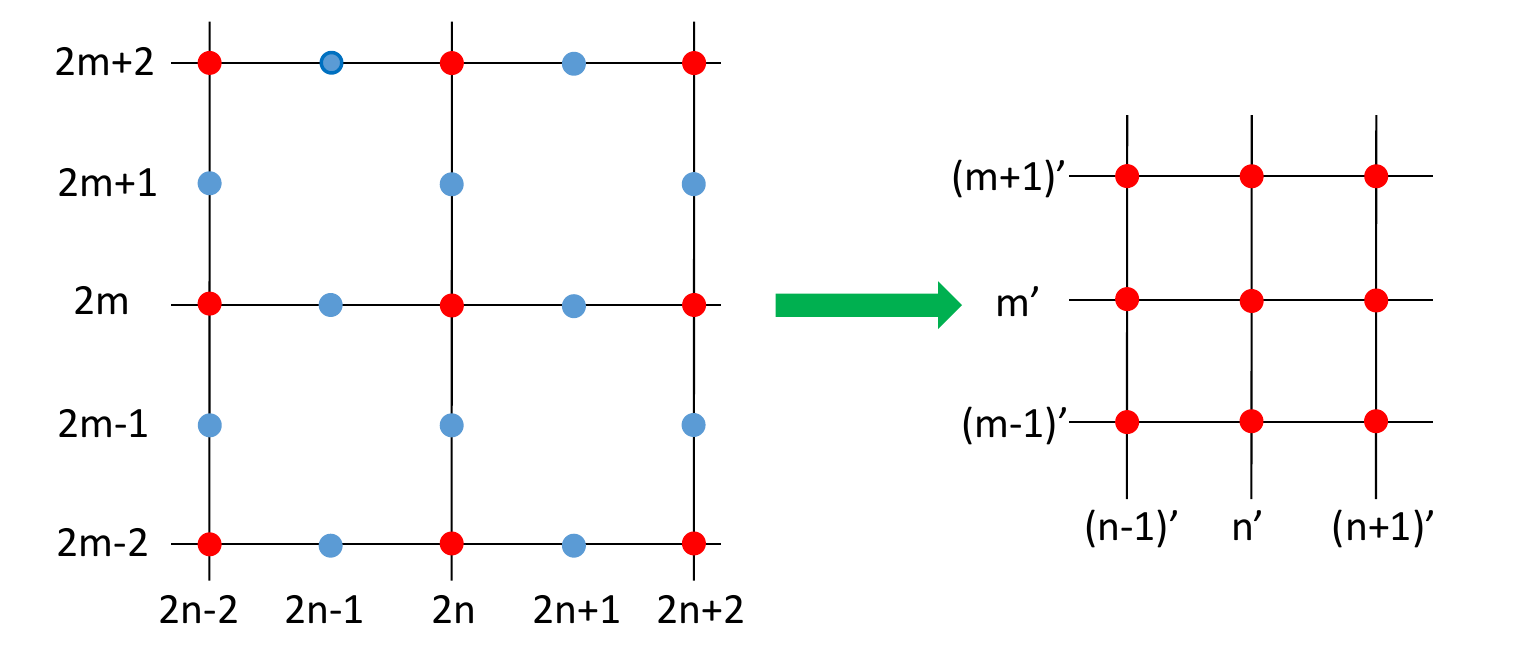}
\caption{\label{S1}
  Mapping 2D VDLL model with $\rho=2$ to 2D AA model. Here the quasiperiodic potentials only act on the vertices (red spheres).}
\end{figure}

\section{II. mapping 2D VDLL model to 2D AA model}
In the main text, we deform the Hamiltonian~(\ref{Hsums}) to 2D AA model. To illustrate this process more concretely, in this section, we show the details of the mapping from 2D VDLL model with $\rho=2$ to 2D AA model.
Suppose that a eigenstate is described by $|\psi\rangle=\sum_{i,j}u_{i,j}c^{\dagger}_{i,j}|0\rangle$, by using the eigenvalue equation $H|\psi\rangle=E|\psi\rangle$, we have
\begin{equation}\label{hamS1}
Eu_{2m,2n} = u_{2m+1,2n}+u_{2m-1,2n}+u_{2m,2n+1}+u_{2m,2n-1}+V_{2m,2n}u_{2m,2n},
\end{equation}
and
\begin{eqnarray}\label{hamS2}
Eu_{2m+1,2n} &=& u_{2m+2,2n}+u_{2m,2n},\qquad Eu_{2m-1,2n} = u_{2m,2n}+u_{2m-2,2n},\nonumber\\
Eu_{2m,2n+1} &=& u_{2m,2n+2}+u_{2m,2n},\qquad Eu_{2m,2n-1} = u_{2m,2n}+u_{2m,2n-2},
\end{eqnarray}
By using Eq.~(\ref{hamS2}), $u_{2m\pm 1,2n}$ and $u_{2m,2n\pm 1}$ in Eq.~(\ref{hamS1}) can be replaced, and then, Eq.~(\ref{hamS1}) becomes
\begin{eqnarray}\label{ha}
(E^2-4)u_{2m,2n} &=& u_{2m+2,2n}+u_{2m-2,2n}+u_{2m,2n+2}+u_{2m,2n-2}+EV_{2m,2n}u_{2m,2n},\nonumber\\
&=& u_{2(m+1),2n}+u_{2(m-1),2n}+u_{2m,2(n+1)}+u_{2m,2(n-1)}+EV_{2m,2n}u_{2m,2n}.
\end{eqnarray}
The indexes are divided by $2$, i.e., $2m\rightarrow m', 2n\rightarrow n', 2(m\pm1)\rightarrow (m\pm1)', 2(n\pm1)\rightarrow (n\pm1)'$ and $E^2-4\rightarrow E'$, as shown in Fig.~\ref{S1}.  Then the above equation map to the 2D AA model with the effective quasiperiodic potential $V'_{m',n'}=2V'[\cos(2\pi\alpha_1m')+\cos(2\pi\alpha_2 n')]$, where $V'=EV$, so the extended-localized transition point is at
$|V'|=1\rightarrow|VE|=1$, and the localization length $\xi=\frac{1}{\ln(|V'|)}\rightarrow\xi=\frac{2}{\ln(|VE|)}$ after considering that the system size need to be enlarged 2 times when mapping the 2D AA model back to the VDLL model.

\begin{figure}[t]
\includegraphics[width=0.6\textwidth]{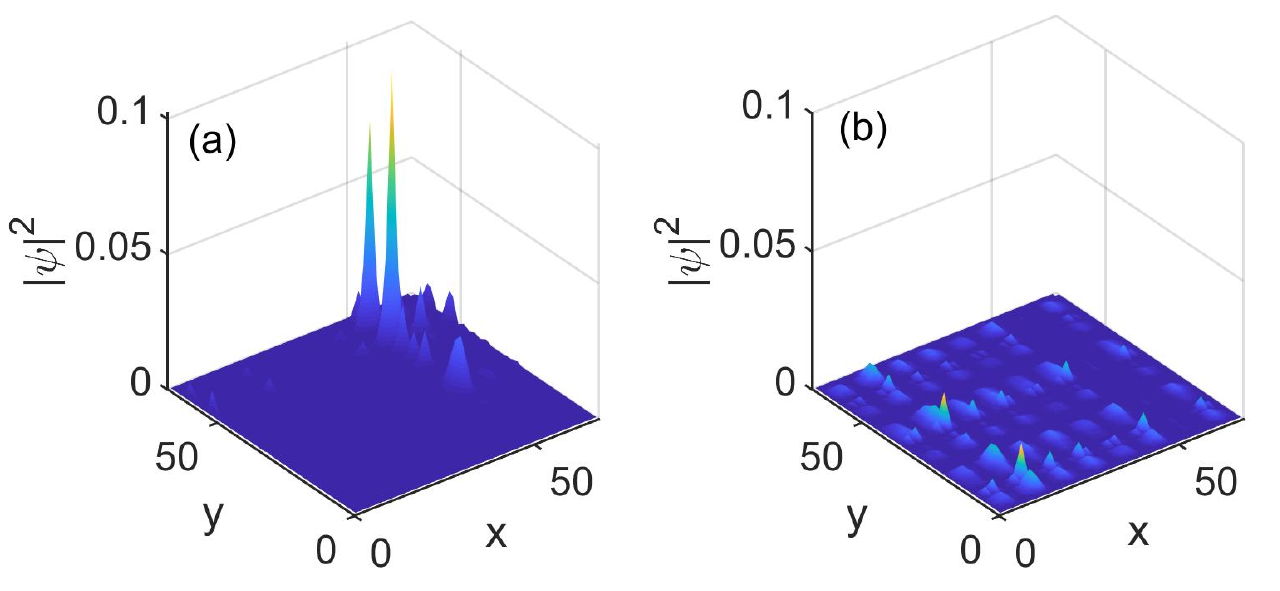}
\caption{\label{S1s}
  Spatial distributions of two eigenstates with the corresponding eigenvalues (a) $E=-2.167(7)$ and (b) $E=-1.920(0)$, which are respectively below and above the ME of the VDLL model with $V=0.5$ and size $L_x=L_y=34$.}
\end{figure}

The mobility edge (ME) can be further confirmed by computing the spatial distributions of wave functions, as shown in
Fig.~\ref{S1s}. The wave functions for $\rho = 2$ are extended
and localized when their eigenvalues satisfy $|E|<1/V$ and $|E|>1/V$, respectively.

\section{III. The underlying mechanism for the occurrence of robust flat bands}
\begin{figure}[h]
\includegraphics[width=0.55\textwidth]{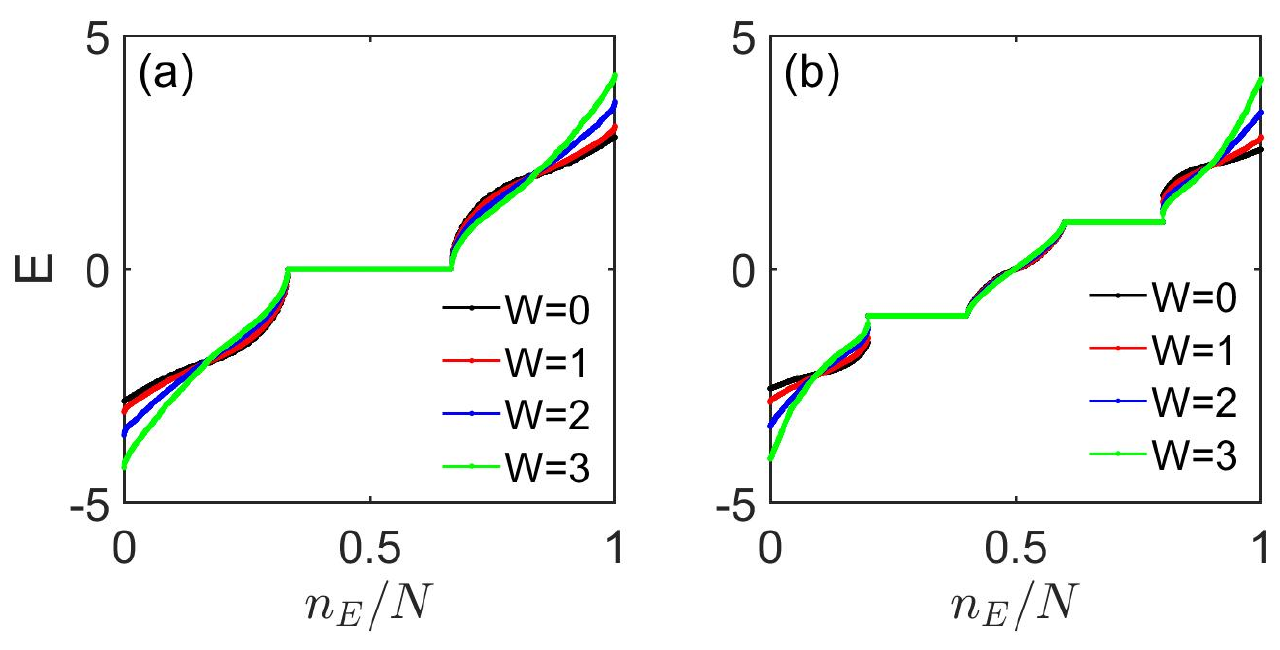}
\caption{\label{S2}
  Eigenvalues as a function of $n_E/N$ for (a) the $\rho=2$ (Lieb lattice) case and (b) the $\rho=3$ (extended Lieb lattice) case with different disorder strength $W$. The system size is $L_x=L_y=34$.}
\end{figure}

In this section, we firstly investigate the effect of random disorder on the flat bands. Here random disorder only act on the vertexs (red spheres in Fig.1 in the main text) of the Lieb lattice or extended Lieb lattices, i.e., $V_{i,j}\in[-W,W]$ when $i=m\rho$ and $j=n\rho$. Fig.~\ref{S2}(a) and (b) show the eigenvalues as a function of $n_E/N$ for the $\rho=2$ and $\rho=3$, respectively, and here $n_E$ is the index of eigen-energy.
With increasing the strength of random disorder potentials, the number of eigenstates in flat bands remain unaffected. The results of the level-spacing ratio are also similar to Fig.3(c) and (d) in the main text, suggesting that these states in the flat band are localized.
These results are not unexpected as the discussions in the main text. 

Then we discuss the underlying mechanism for the occurrence of the robust flat bands. In the absence of the on site potentials, i.e., $V_{ij}=0$ in Eq.~\ref{Hsums}, Fourier transforming this Hamiltonian results in the Hamiltonian in the momentum space, which is a $3\times 3$ matrix because of the three inequivalent lattice sites per unit cell [see Fig.1(a) in the main text or Fig.~\ref{S4}(a)],
\begin{equation*}
H=\sum_{k}(c^{\dagger}_{k,A}, c^{\dagger}_{k,B}, c^{\dagger}_{k,C})h(k)\left(
\begin{array}{c}
  c_{k,A} \\
  c_{k,B} \\
  c_{k,C} \\
\end{array}
\right)
\end{equation*}
where
\begin{equation*}
h(k)=\left(
\begin{array}{ccc}
  0& 2J\cos(k_y)& 2J\cos(k_x) \\
  2J\cos(k_y) & 0 & 0 \\
  2J\cos(k_x) & 0 & 0
\end{array}
\right),
\end{equation*}
which has the simple form
\begin{equation}\label{hkSk}
h(k)=\left(
\begin{array}{cc}
  0& S_{k} \\
  S^{\dagger}_{k} & 0 \\
\end{array}
\right).
\end{equation}
$S_{k}$ is a $1\times 2$ matrix, and we can write the singular value decomposition of $S_{k}$ as
\begin{equation}\label{SVD1}
S_{k}=V_k\Sigma_kD_k=\sum^{r_k}_{\alpha=1}\epsilon_{k,\alpha}\phi_{k,\alpha}\psi^{\dagger}_{k,\alpha},
\end{equation}
where $V_k$ ($D_k$) is a $1\times1$ ($2\times2$) unitary matrix whose columns form eigenstates of $S_kS^{\dagger}_{k}$ ($S^{\dagger}_{k}S_k$), and $r_k$ is the rank of $S_k$. Due to that $S_{k}$ is a $1\times 2$ matrix, we have $r_k=1$. $\Sigma_k$ is a $1\times2$ matrix: $(\epsilon_{k}, 0)$ with $\epsilon_{k}$ being the singular value of $S_k$. The $\alpha-$th column of $V_k$ ($D_k$) is denoted by $\phi_{k,\alpha}$ ($\psi_{k,\alpha}$) being the $\alpha-$th left (right) singular eigenvector of $S_k$. Here $\alpha$ has to equal to $1$, so $\phi_k=V_k$ and $\psi_k$ is
the first column of $D_k$.
By using Eq.~\ref{SVD1}, we perform a unitary transformation of $h(k)$
\begin{equation*}
h(k)=\left(
\begin{array}{cc}
  V_k& 0 \\
  0 & D_k \\
\end{array}
\right)\left(
\begin{array}{cc}
  0 & \Sigma_k \\
  \Sigma^{T}_k & 0 \\
\end{array}
\right)\left(
\begin{array}{cc}
  V^{\dagger}_k& 0 \\
  0 & D^{\dagger}_k \\
\end{array}
\right).
\end{equation*}
Due to $\Sigma_k=(\epsilon_{k}, 0)$, $h(k)$ is similar to a matrix containing one zero row and column, which induce that $h(k)$ has at least one zero mode for any $k$~\cite{Bernevig2022}. Consequently, Lieb lattice has one flat band pinned at zero energy.

From the above discussions, if $S_k$ in Eq.~\ref{hkSk} is a $N_1\times N_2$ matrix, $h(k)$ will necessarily has at least $|N_1-N_2|$ zero modes at each momentum point, suggesting that this system will feature at least $|N_1-N_2|$ flat bands~\cite{Bernevig2022}, which can also be clearly seen from the real space. When quasiperiodic or random disorder potentials are added on vertices of the Lieb lattice, $S_k$ is unaffected, and so the flat band is robust. When the potentials are added on the edges, or existing hopping between the lattice site B and lattice site C, $h(k)$ can not be written as Eq.~\ref{hkSk}, and then, the flat bands will be easily destroyed.

\section{IV. Experimental detection and realization}
\begin{figure}[t]
\centering
\includegraphics[width=0.65\textwidth]{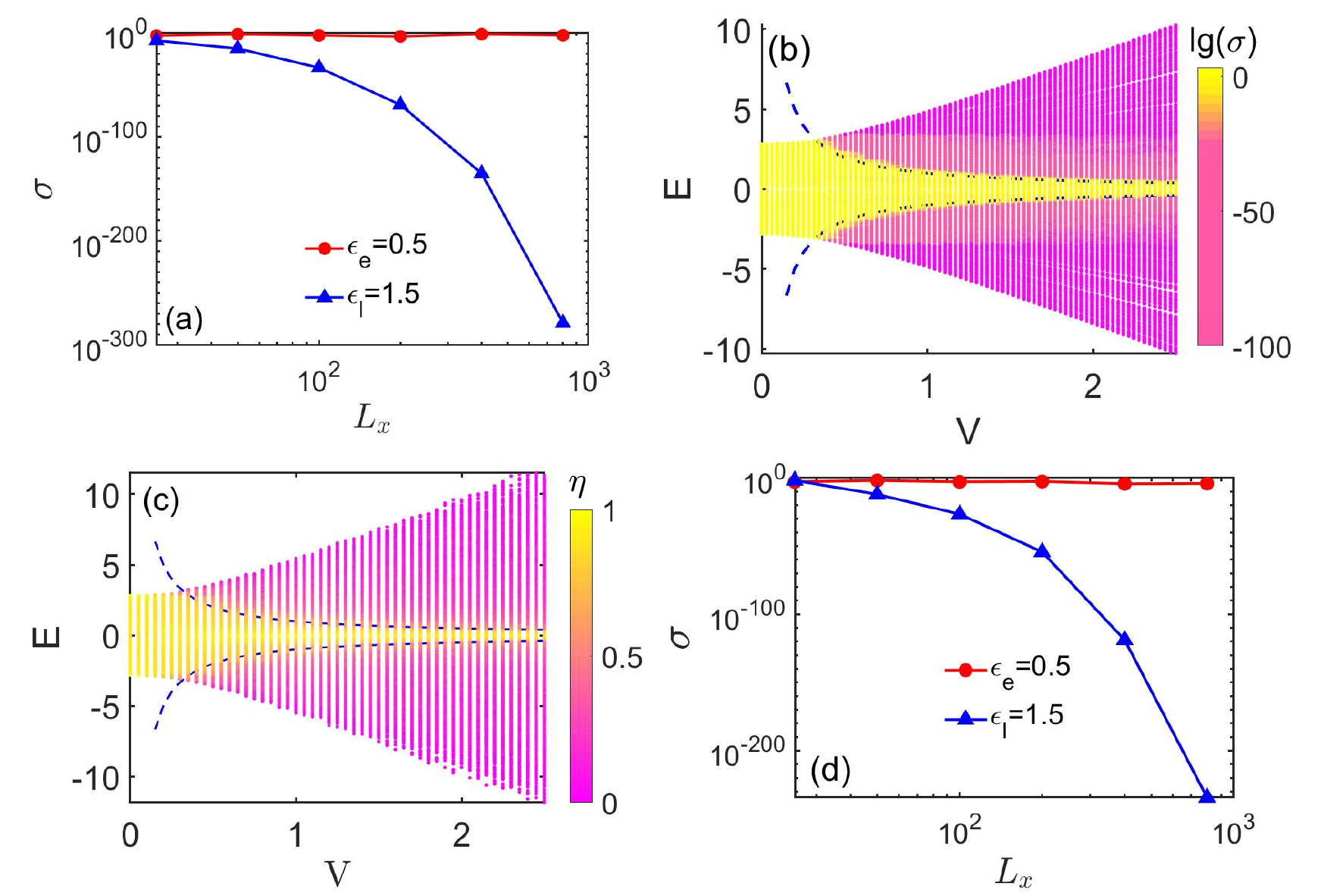}
\caption{\label{S3}
(a) $\sigma$ versus $L_x$ with fixed $L_y=100$ and $V=1$. The Fermi surfaces in the extended and localized regions are chosen at $\epsilon_e=0.5$ and $\epsilon_l=1.5$, respectively. (b) Conductance of different eigenvalues as a function of the corresponding eigenvalues $E$ and the quasiperiodic potential strength $V$.  Blue dashed lines represent the MEs predicted by theory. The system size is $L_x=L_y=25$. (c) Fractal dimension as a function of $V$ and eigenvalues $E$ for the VDLL model with $\rho=2$ and the system size $L_x=L_y=34$, and in this system, $V_{ij}$ is replaced by $V_{ij}+0.2V_{ij}R_{ij}$ with $R_{ij}\in[-1,1]$ being random numbers. (d) $\sigma$ versus $L_x$ when adding the term $0.2V_{ij}R_{ij}$, and other parameters are same with Fig.(a).
}
\end{figure}

\begin{figure}[h]
\centering
\includegraphics[width=0.8\textwidth]{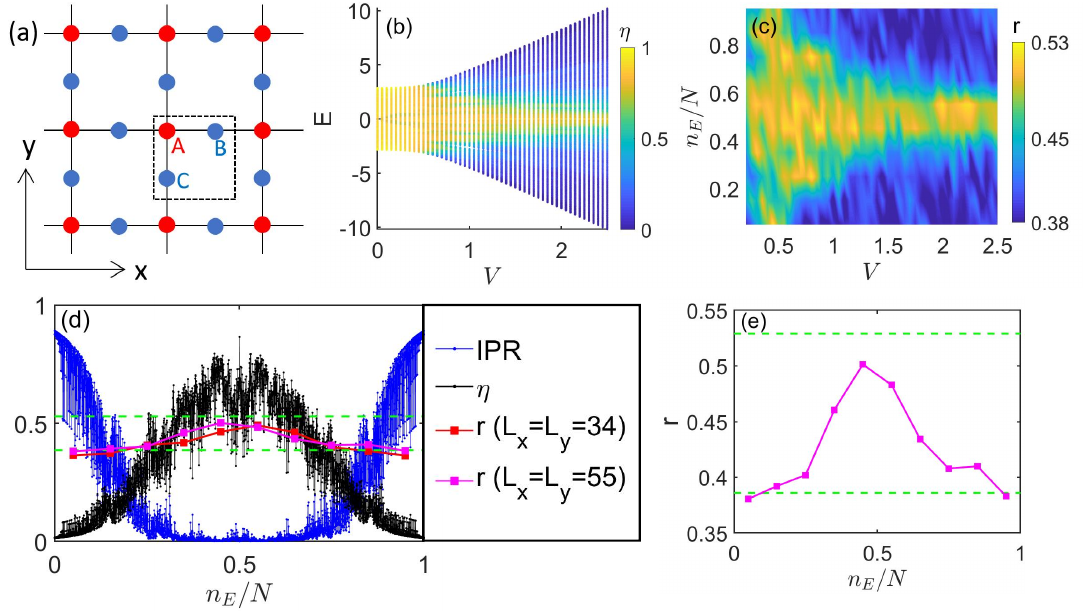}
\caption{\label{S4}
(a) Schematic figures of the Lieb lattice, and the quasiperiodic potentials only act on the edges (blue spheres). (b) Fractal dimension $\eta$ as a function of $V$ and corresponding eigenvalues $E$ of the edge-decorated Lieb lattice system with the system size $L_x=L_y=34$. (c) $\langle r\rangle$ versus $V$ and different energy windows $n_E/N$. (d) IPR and $\eta$ of different eigenstates with fixed $V=1.5$ and $L_x=L_y=34$, and $\langle r\rangle$ of different energy windows for different sizes. Green dashed lines correspond to $0.529$ and $0.387$, respectively. (e) The enlarged part of $\langle r\rangle$ with $L_x=L_y=55$.}
\end{figure}

For the quantum dot system in the main text, the predicted MEs could be detected by detecting the transport signal.
The transport properites of the system with quasiperiodic potential are investigated by using the non-equilibrium Green's function method \cite{Keldysh1965}and Landauer-B\"{u}ttiker formula~\cite{Landauer1970,Fisher1981,Buttiker1988}.  The conductance can be written as
\begin{equation}\label{cond}
\sigma\left( \varepsilon  \right) = \frac{{{e^2}}}{h}T\left( \varepsilon  \right),
\end{equation}
where $T\left( \varepsilon  \right) = Tr[\Gamma _{left}^{}\left( \varepsilon  \right){G^R}\left( \varepsilon  \right)\Gamma _{right}^{}\left( {\varepsilon )} \right){G^A}\left( \varepsilon  \right)]$ is the transmission coefficient at energy $\varepsilon$. The linewidth function $\Gamma _{left/right}^{} = i[\Sigma _{left/right}^R - \Sigma _{left/right}^A]$, and the Green's functions ${G^{R/A}}\left( \varepsilon  \right)$ can be obtained by ${G^R}\left( \varepsilon  \right) = {[{G^A}\left( \varepsilon  \right)]^\dagger } = {[\varepsilon I - H_c - \Sigma _{left}^R - \Sigma _{right}^R]^{ - 1}}$
, where $H_c$ is the Hamiltonian of the central scattering region and $\Sigma _{left/right}^{R/A}$ are the retarded (advance) self-energies due to the attatching of the left(right) lead. By using the non-equilibrium Green's function method, we numerically calculate the conduction of the 2D VDLL models, which is adopted as the central scattering device in the process of calculation, as shown in Figs.~\ref{S3}(a) and (b). One can also detect the scaling behaviors of the conductance to distinguish the extended states from localized ones. In the extended regions, the conductivity is independent of the system sizes [red data points in Fig.~\ref{S3}(a)], while in the localized regions, conductivity decays exponentially with the system sizes [blue data points in Fig.~\ref{S3}(a)]~\cite{Archak2017,Archak2019}. Since the conductivity shows very fast size-dependent attenuations in the localized region, when the system size is large enough, the conductance $\sigma$ suddenly changes when energies across the MEs, as shown in Fig.~\ref{S3}(b).

To investigate the errors caused by the small inaccuracies of $V_{ij}$, in Fig.~\ref{S3}(c) and (d), we respectively show the fractal dimensions and the scaling behaviors of the conductance after that the quasiperiodic potential $V_{ij}$ is replaced by $V_{ij}+0.2V_{ij}R_{ij}$ with $R_{ij}\in[-1,1]$ being random numbers. We see that the position of MEs and the corresponding transport properties in extended and localized regions are scarcely influenced, so the experimental realization of $V_{ij}$ is of high fault tolerance.

\section{V. Critical regions in 2D systems}
In the main text, we have considered the VDLL models. In this section, we consider the other case that the quasiperiodic potentials
only act on the edge sites [blue spheres in Fig.~\ref{S4}(a)], i.e., $V_{ij}$ in the Hamiltonian~(\ref{Hsums}) becomes
\begin{equation}\notag
V_{ij}=
\begin{cases}
0,\ \  (i,j)=(m\rho,n\rho)  \\
2V[\cos(2\pi\alpha_1 i+\theta_1)+\cos(2\pi\alpha_2 j+\theta_2)], \ \textrm{otherwise},
\end{cases}
\end{equation}
Fig.~\ref{S4}(b) shows the fractal dimensions $\eta$, which is defined as $\eta=-\lim_{N\rightarrow\infty}\ln(IPR)/\ln N$, where $IPR=\sum_{ij}u^4_{ij}$ is the inverse participation ratio (IPR) and $N$ is the number of the total lattice sites. The fractal dimension tends to $0$, $1$ and $0<\eta<1$ for the localized, extended and critical states, respectively. It can be see that there exist different regions, in which the localization properties are different. To further see the localized properties of different regions, we list the eigenvalues in ascending order and then divide the total levels into ten parts, meaning that every part has $N/10$ eigenvalues. As shown in the main text, we can calculate the level-spacing ratio $r_k=\frac{min(\delta_k,\delta_{k+1})}{max(\delta_k,\delta_{k+1})}$, where $\delta_k=E_{k+1}-E_k$ is the energy spacing, and then obtain the averaged ratio $\langle r\rangle$ for every part. In the localized region, the spectral statistics are Poisson, which yields $\langle r\rangle\approx 0.387$. In the extended region, the spectral statistics follow Gaussian-orthogonal ensemble (GOE) yielding $\langle r\rangle\approx 0.529$. In the critical region, the spectral statistics are neither Poisson nor GOE, but are well described by the critical statistics, which induces that $\langle r\rangle$ is neither 0.387 nor 0.529. Fig.~\ref{S4}(c) displays
the $\langle r\rangle$ of different energy windows $n_E/N$. The values $\langle r\rangle$ at the position $n_E/N=0.5\kappa$ correspond to that the average $r$ is taken in the region $n_E/N\in [(\kappa-1)N/10, \kappa N/10]$. We see that besides the values approaching $0.529$ and $0.387$, there also exist the $\langle r\rangle$ being not close to the two values, manifesting a different region from extended and localized regions.  To further confirm this, we present a quantitative study with $V=1.5$. Fig.~\ref{S4}(d) shows the $IPR$ and $\eta$ of different eigenstates, and $\langle r\rangle$ of different energy windows. We see that $IPR$ gradually increase and $\eta$ gradually decrease from the center to either side of the energy spectra, suggesting that eigenstates gradually change from delocalization to localization.
Generally, $\eta$ for extended and localized states are obviously different, and there should exist sudden change at MEs~\cite{Wang2022}.
Further, $\langle r\rangle$ in band tails are close to $0.387$, suggesting that the corresponding states are localized, while in the center, $\langle r\rangle$ is not near $0.387$ or $0.529$ [see Fig.~\ref{S4}(e)], suggesting that the corresponding states are critical. Thus, there should exist MEs separating the localized states from critical ones. For other $V$, there should also exist MEs separating the extended states from critical ones.

\section{VI. Quasiperiodic potentials act on all the Lieb lattice sites}
\begin{figure}[t]
\centering
\includegraphics[width=0.8\textwidth]{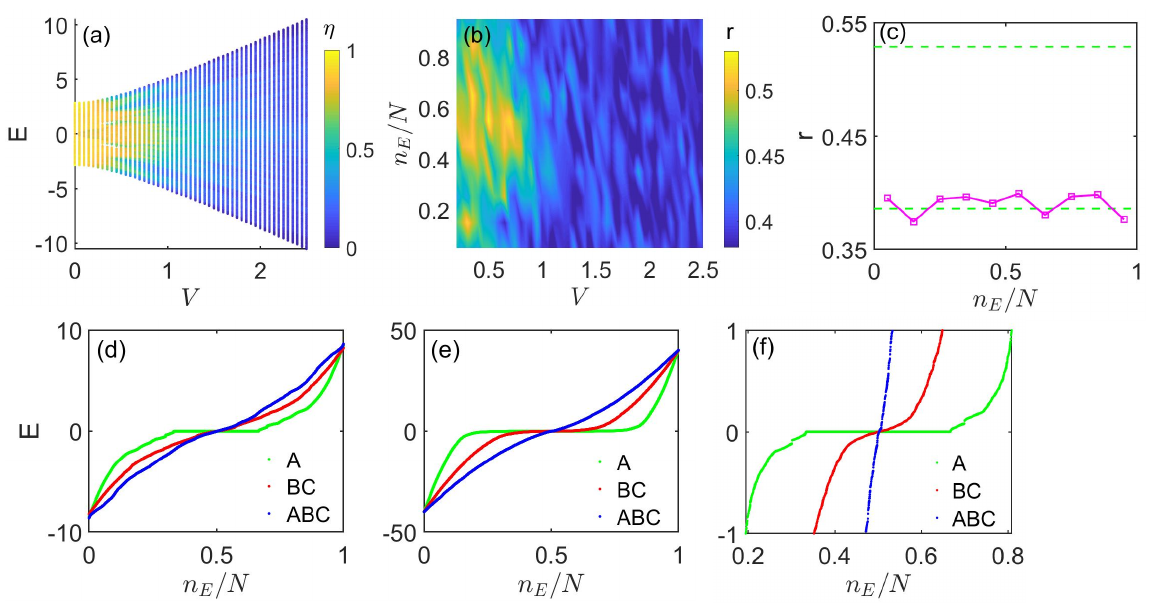}
\caption{\label{S5}
(a) Fractal dimension $\eta$ as a function of quasiperiodic potential strength $V$ and eigenvalues $E$ of the system with the system size $L_x=L_y=34$. (b) $\langle r\rangle$ versus $V$ and energy windows $n_E/N$. (c) $\langle r\rangle$  of different energy windows for the system with with $L_x=L_y=55$. Green dashed lines respectively correspond to $0.529$ and $0.387$. Eigenvalues as a function of $n_E/N$ with $n_E$ being the index of eigen-energies with (d) $V=2$ and (e) $V=10$. (f) The enlarged part of the energy window [-1,1] in Fig.(e). For (d)(e)(f), the green, red, and blue data points correspond to that the quasiperiodic potentials act on the vertices (red spheres in Fig.~\ref{S4}(a)), edges (blue spheres Fig.~\ref{S4}(a)), and all sites respectively, and we fix $L_x=L_y=34$.}
\end{figure}

In this section, we consider the case that the quasiperiodic potentials act on all the Lieb lattice sites, i.e., $V_{ij}$ in the Hamiltonian~(\ref{Hsums}) becomes $V_{ij}=2V[\cos(2\pi\alpha_1 i+\theta_1)+\cos(2\pi\alpha_2 j+\theta_2)]$.
Fig.~\ref{S5}(a) shows the fractal dimensions $\eta$, and we can see that there also exist MEs. Comparing the MEs in the three cases that quasiperiodic potentials act on the vertices [Fig.~\ref{S3}], edges [Fig.~\ref{S4}(a)], and all sites [Fig.~\ref{S5}(a)], one can find that for the first two cases, the states near $E=0$ remain delocalized although the quasiperiodic potential strength is large, while for latter, all states become localized when $V>1$. This phenomenon can be further determined by using level-spacing ratio, as shown in Figs.~\ref{S5}(b) and (c) [compare with Figs.~\ref{S4}(c) and (e)]. It can be seen that all $\langle r\rangle$ are close to $0.387$ when $V>1$, suggesting that all states are localized.
As we known, $\langle r\rangle$ reflects the distribution of eigen-energies. We below consider the detailed distributions, and show the eigenvalues as a function of $n_E/N$ in Figs.~\ref{S5}(d), (e) and (f). Comparing the two systems with $V=2$ [Fig.~\ref{S5}(d)] and $V=10$ [Fig.~\ref{S5}(e)], the number of states near $E=0$ in the latter is significantly increased when quasiperiodic potentials only act on the vertices (green data points) and edges (red data points). Fig.~\ref{S5}(f) is the enlarged part of the energy window [-1, 1] in Fig.~\ref{S5}(e), and we see that the density of states near $E=0$ is increased and the number of states in the flat band remains unchanged, meaning that the increased states do not get into the flat band. Since the large density of states is against localization, the states near $E=0$ remain delocalized even when the quasiperiodic strength is large. When quasiperiodic potentials act on all sites, comparing the blue data points in Figs.~\ref{S5}(d) and (e), one can see that
the number of states near $E=0$ is not obviously changed. Thus, the states near $E=0$ in this system easily become localized compared with the first two cases.

\begin{thebibliography}{36}
\bibitem{explainloc} Without loss of generality, we take $\rho=2$ as an example and set $EV/J^2>0$. From the expression of the localization lengths, we have $\xi(E)=\frac{\rho}{\ln|EV/J^2|}\approx\frac{2}{EV/J^2-1}=2J^2/V(E-J^2/V)^{-1}$ when fix the quasiperiodic potential strength $V$, and when fix $E$, we have $\xi(E)\approx 2J^2/E(V-J^2/E)^{-1}$. We introduce a critical exponent $\nu$, and then the localization length can be written as $\xi\sim(V-V_c)^{-\nu}$ or $\xi\sim(E-E_c)^{-\nu}$, where $E_c=J^2/V$ and $V_c=J^2/E$ are respectively the critical energy and critical quasiperiodic potential strength with fixing $V$ and $E$ [see Eq.~(\ref{rho2})]. Thus, we have the critical exponent $\nu=1$, which possess a wide suitability for other 2D systems.
\bibitem{explain2} The Gaussian-orthogonal ensemble data in Fig.~\ref{03}(c) was obtained from $500$ random matrices of size $4000$.
\bibitem{explain3} In the process of calculating $P(r)$ and $\langle r\rangle$, we did not take different samples, because for non-interacting quasiperiodic systems, many properties are independent of the phase shifts~\cite{explain4}. Thus, if considering different samples by choosing the initial phases $\theta_1$ or $\theta_2$, the results do not have any difference.
\end{thebibliography}
\end{document}